\documentstyle[11pt]{article}

 \newdimen\paperwidth \newdimen\paperlength \newdimen\margin
\newdimen\vmargin
\textwidth=\paperwidth \advance \textwidth by -2\margin
\advance\hoffset by -1truein \advance\hoffset by \margin
\textheight=\paperlength \advance \textheight by -2\vmargin
\advance\voffset by -1truein \advance\voffset by \vmargin
\advance\textheight by -2\baselineskip
\begin{document}

\renewcommand{\theequation}{\thesection.\arabic{equation}}
\newcommand{\Section}[1]{\section{#1}\setcounter{equation}{0}}

 \begin{titlepage} \title{ {\bf   The Renormalization  Group
Method \\ and Quantum Groups:} \\ {\bf the postman always rings twice
}\thanks{Work partly supported by CICYT under contracts
AEN93-0776 (M.A.M.-D.) and PB92-109 ,
 European Community Grant ERBCHRXCT920069 (G.S.).} }

\vspace{2cm}    \author{ {\bf Miguel A. Mart\'{\i}n-Delgado}\dag
\mbox{$\:$} and {\bf Germ\'an Sierra}\ddag \\ \mbox{}    \\ \dag{\em
Departamento de F\'{\i}sica Te\'orica I}\\ {\em Universidad
Complutense.  28040-Madrid, Spain }\\ \ddag{\em Instituto de
Matem\'aticas y F\'{\i}sica Fundamental. C.S.I.C.}\\ {\em Serrano 123,
28006-Madrid, Spain } } \vspace{5cm}
 \date{} \maketitle \def\baselinestretch{1.3} \begin{abstract}

We review some of our recent
 results concerning the relationship between the
Real-Space Renormalization Group method and Quantum Groups.
We show this relation by applying real-space RG methods to study
two quantum group invariant Hamiltonians, that of the XXZ model
and the Ising model in a transverse field (ITF)
defined in an open chain
with appropriate boundary terms. The quantum group symmetry is
preserved under the RG transformation except for the appearence of a
quantum group anomalous term which vanishes in the classical case.
This is called {\em the quantum group anomaly}.
 We derive the new qRG equations for the XXZ model
 and show that the RG-flow diagram obtained in this fashion exhibits
the correct line of critical points that the exact model has.
In the ITF model the qRG-flow equations
coincide with the tensor product decomposition of cyclic irreps of
$SU_q(2)$ with $q^4=1$.

\end{abstract}

\vspace{2cm} {\em TO HONOR JERZY IN THIS CELEBRATED DATE}

\vskip-17.0cm \rightline{UCM/CSIC-95-10} \rightline{{\bf October
1995}} \vskip3in \end{titlepage}

\newpage

\Section{Introduction}

The Renormalization Group method has become one of the basic concepts
in Physics, ranging from areas such as Quantum Field Theory and
Statistical Mechanics to Condensed Matter Physics. The many interesting
and relevant models encountered in these fields are usually not
exactly  solvable except for some privileged cases in one dimension.
It is then when we resort to the RG method to retrieve the essential
features of those systems in order to have a qualitative understanding
of what the physics of the model is all about. This understanding is
usually recasted in the  form of a RG-flow diagram were the different
possible behaviours of  the model leap to the eyes.

Many authors in the past have contributed significantly to the idea of
renormalization and it is out of the scope of this paper to give a
detailed account on this issue here. We shall be dealing with the the
version of the RG as introduced by Wilson
 \cite{wilson} and Anderson \cite{anderson}
 in their treatment of the Kondo problem, and subsequent  developments
of these ideas carried out by Drell et al. at the SLAC group
\cite{drell} and Pfeuty et al. \cite{jullien}.

It was Wilson in the late sixties and early seventies
 who set up the framewok of the method in its more thorough and
complete version. According to his own words,  he did so in his
search for a better understanding of what a  quantum field theory is.
To this end he also introduced another tool, the Operator  Product
Expansion (OPE) for the field operators. Both the RG method and the
OPE have become two conerstones in modern Quantum Field Theory.

\noindent It has been long known that the physics of (1+
1)-dimensional quantum many-body systems and 2D statistical field
theories were very special in many aspects when compared to their
higher dimensional generalizations.  It was after the seminal work by
Belavin, Polyakov and Zamolodchikov (BPZ) \cite{bpz}
 when the special role of conformal symmetry in two dimensions was
brought  about in connection to those many special properties
exhibited by 2D systems such as, criticality, integrability, etc...
The basic tool employed by BPZ to develop their conformal program was
precisely one of the tools introduced by Wilson, the OPE, which with
the help of  two-dimensional conformal symmetry is powerful enough so
as to classify the  local fields forming the local algebra according
to the irreducible representations of the Virasoro algebra and to
determine the correlation functions of those fields.

\noindent Among all the Conformal Field Theories studied by BPZ they
singled out what they called Minimal Models as those for which the
conformal program is more successful in determining their properties
the best. The minimal models, also called Rational Conformal Field
Theories (RCFT), are those CFT which contain finite number of primary
fields in the operator algebra. For these models the anomalous
dimensions of the operators, or equivalently the critical exponents,
are known exactly and moreover, their correlations functions can be
computed as solutions of special systems of linear differential
equations. The underlying phenomenon responsible for such remarkable
features is the  truncation of the operator algebra, that is, the
primary fields form a closed  operator algebra.

After the introduction of Quantum Groups by Drinfeld
\cite{drinfeld-jimbo}
in the mid-eighties, some CFT theorists realized the relationship
between these new algebraic structures and those appearing in the
RCFT. In reference \cite{ags} it was shown that the representation
theory of $SL(2,q)$ with $q$ a root of unity provides solutions to the
polynomial equations of RCFT, as well as a rather efficient way to
compute the duality matrices. The rationality condition is met by
requiring $q$ to be a root of unity. In this case the representation
theory of the quantum group is significantly different from the
classical one. The first surprise is that when $q^N=1$ there are only
$N-1$ distinct finite dimensional irreps with spins
$j=0,1/2,1,\ldots ,(N-1)/2$, this last one irrep being singular.
This truncation of the representation theory of quantum groups
was put in correspondence \cite{ags} with the previous truncation
of the operator algebra found by BPZ in the minimal models.

With this historical perspective in mind, what we have suggested
by introducing what we call the Renormalization Quantum Group Method
(qRG for short) \cite{q-germanyo}
 is to establish the connection of the truncation of
 states characteristic of the Real-Space RG methods (the other tool
introduced by Wilson) with the special features of the
(1+1)-dimensional systems exemplified by the RCFT.
The connection we have found can be casted into the following
squematical form,

\[
 \mbox{$q$RG-truncation} \leftrightarrow \mbox{RCFT}
\]

\noindent whose precise content will be explained in the following
sections with several examples.

 Physicists working in field theory and condensed  matter generalized
the Real Space Renormalization Group methods introduced by  Wilson
\cite{wilson} to other problems by using the Kadanoff's concept of
block \cite{drell}, \cite{jullien}. The Block method (BRG) has the
advantage of  being conceptually and technically simple, but it lacks
of numerical accuracy  or may even produce wrong results. For this
reason the analytical BRG methods  were largely abandoned in the 80's
in favor of numerical methods such as the  Quantum Monte Carlo
approaches. In the last few years there has been new  developments in
the numerical RG methods motivated by a better understanding of the
errors introduced by the splitting of the lattice into disconnected
blocks. A  first step was put forward in \cite{white-noack} where a
combination of different  periodic boundary conditions applied to
every block lead to the correct energy  levels of a simple
tight-binding model. This method however has not been generalized  to
models describing interactions.  Recently, we have clarified the role
played by the boundary conditions in the real-space  renormalization
group method \cite{bc-germanyo} by constructing a new analytical
BRG-method which is able to give the exact ground state  of the model
and the correct $1/N^2$-law for the energy of the first excited state
in the  large $N$(size)-limit. A further step in the generalization to
interacting models was  undertaken by White in \cite{white} where a
Density Matrix algorithm (DMRG) is  developed. The main idea is to
take into account the connection of every block with the  rest of the
system when choosing the states which survive the truncation
procedure. The standard prescription is to choose the lowest energy
states of the block Hamiltonian. Instead, in the DMRG method one
replaces the block  Hamiltonian by a block density matrix and chooses
the eigenstates of this matrix with the  highest eigenvalues. The
density matrix is constructed out of the ground state of  a superblock
which contains the desired block.

In these notes we want to explain another RG method, the qRG method,
 which uses the  concept of quantum groups. This mathematical notion
emerged in the study of  integrable systems and it has been applied to
conformal field theory, invariants of  knots and manifolds, etc.
\cite{drinfeld-jimbo}, \cite{germancesar}. The new application of
quantum groups that we envisage has been partially  motivated by the
aforementioned work of White, Noack and collaborators and it is
probably related to it. This relation is suggested by the fact that
quantum groups describe symmetries in the presence of non-trivial
boundary conditions.  The typical example to understand this property
of quantum groups is given by  the 1D Heisenberg-Ising model with
anisotropic parameter $\Delta $. The  isotropic model $\Delta = \pm1$
is invariant under the rotation group $SU(2)$, but as  long as
$|\Delta| \neq 1$ this symmetry is broken down to the rotation group
$U(1)$ around the z-axis. One can ``restore" this full rotation
symmetry by  adding appropriate boundary operators to the Hamiltonian
of the open chain. The  classical group $SU(2)$  becomes then the
quantum group $SU_q(2)$, where the  quantum parameter is related to
the anisotropy by $\Delta  = \frac{q+q^{-1}}{2}$ \cite{sklyanin},
\cite{pasquier-saleur}. The ``restoration" of a  classical symmetry
into a $q$-symmetry is achieved at the price of deforming the  algebra
and the corresponding addition rule of angular momentum. The $q$-sum
rule, which is called the comultiplication, becomes non local and
violates parity.  The total raising (lowering) operators acting on the
whole chain are a sum of the  raising ( lowering) operators acting at
every single site times  a non-local term  involving all the remaining
sites which appear in an asymmetric way: sites located  to the left or
to the right at a given site contribute differently.

These features of $q$-groups made them specially well-suited to
implement  a RG method which takes into account the correlation
between neighboring  blocks. In the forthcoming sections we show how
this can be done explicitly in two examples in 1D: the
Heisenberg-Ising model and the Ising model in a transverse field (ITF).

\noindent This paper is organized as follows. In Sect.2 we present a
brief introduction to the Block Renormalization Group methods based
upon the concept of  the intertwiner operator $T$. This will allow us
to see the truncation procedure inherent to the BRG method as
tensoring representations of the Hamiltonian symmetry algebra, the
intertwiner operator $T$ being a Clebsch-Gordan operator.
In Sect.3 we use the Heisenberg-Ising model to show how the truncation
process pertaining to the real-space RG is nothing but a tensor
product decomposition of irreps of the symmetry algebra.
The properties of the model are qualitative and quantitative well
described by the BRG equations in the massive region $\Delta > 1$,
while in the massless region one predicts the massless spectrum
{\em but not criticality at each value of } $\Delta > 1$. This latter
fact is rather subtle and elusive.
In Sect. 4 we set up the foundations of the qRG method using the
Heisenberg-Ising model as an example. We derive the new qRG equations
 and show that the RG-flow diagram obtained in this fashion exhibits
the correct line of critical points that the exact model has.
Moreover, the qRG equations for the renormalized spin operators
show the appearence of a novel feature:
{\em the quantum group anomaly}.
In Sect. 5 we apply the qRG method to another model, the Ising model
in a transverse field (ITF model). Here, the qRG-flow equations
coincide with the tensor product decomposition of cyclic irreps of
$SU_q(2)$ with $q^4=1$.
Sect.6 is devoted to
conclusions and prospectives.

\Section{A Brief Review of Block Renormalization Group Methods (BRG)}

In this section the block renormalization group method is revisited
and we present a new  and unified reformulation of it based on the
idea of the {\em intertwiner operator} $T$
 to be discussed  below. For a more extensive account on this method
we refer to \cite{jullienlibro}  and chapter 11 of reference
\cite{jaitisi} and references therein.

The block RG-method is a real-space RG-method introduced and developed
by the SLAC  group \cite{drell}. Let us recall that Wilson developed
his numerical  real-space renormalization group procedure to solve
the  Kondo problem \cite{wilson}.  It was clear from the beginning
that one could not hope to achieve the accuracy  Wilson obtained for
the Kondo problem when dealing with more complicated many-body
quantum Hamiltonians such as Heisenberg, Hubbard, etc ...  The {\em
key difference} is  that in the Kondo model there exists  a {\em
recursion relation} for Hamiltonians at each step  of the
RG-elimination of degrees of freedom.
 The existence of such recursion relation facilitates enormously the
work, but as it  happens it is specific of {\em impurity problems}.

{}From the numerical point of view, the Block Renormalization Group
procedure proved to be not  fully reliable in the past particularly in
comparison with other numerical approaches, such as  the Quantum
MonteCarlo method which were being developed at the same time. This
was one of  the reasons why the BRG methods remained undeveloped
during the '80's until the begining of  the '90's when they are making
a comeback as one of the most powerful numerical tools  when dealing
with zero temperature properties of many-body systems.

Let us first summarize the main features of the real-space RG. The
problem that one faces   generically is that of diagonalizing a
quantum lattice Hamiltonian $H$, i.e.,

\begin{equation} H |\psi > = E |\psi >
\label{1}  \end{equation}

\noindent  where $|\psi >$ is a state in the Hilbert space ${\cal H}$.
If the lattice has N sites  and there are $k$ possible states per site
then the dimension of ${\cal H}$ is simply

\begin{equation} dim{ \cal H} = k^N
\label{2}
 \end{equation}

\noindent As a matter of illustration we cite the following examples:
$k=4$  (Hubbard model), $k=3$ (t-J model), $k=2$ (Heisenberg model)
etc.

\noindent When $N$ is large enough the eigenvalue problem (\ref{1}) is
out of the capability of  any human or computer means unless the model
turns out to be integrable which only happens in some instances in
$d=1$.

\noindent  These facts open the door to a variety of approximate
methods among which the RG-approach is one of the
most relevant. The main idea of the RG-method is the mode elimination
or thinning of the degrees  of freedom followed by an iteration which
reduces the number of variables step by step until  a more manageable
situation is reached. These intuitive ideas give rise to a well
defined  mathematical description of the RG-approach to the low lying
spectrum of quantum lattice  hamiltonians.

To carry out the RG-program it will be useful to introduce the
following objects:

\begin{itemize}

\item ${\cal H}  $ : Hilbert space of the original problem.

\item ${\cal H }' $: Hilbert space of the effective degrees of freedom.

\item $H $: Hamiltonian acting in $\cal H$.

\item $H'$: Hamiltonian acting in $\cal H'$ (effective Hamiltonian).

\item $T  $ : embedding operator : ${\cal H }'\longrightarrow {\cal H}$

\item $T^{\dagger }  $ :truncation operator : ${\cal H}
\longrightarrow {\cal H}'$

\end{itemize}

The problem now is to relate $H$, $H'$ and $T$. The criterium to
accomplish this task is that  $H$ and $H'$ have in common their low
lying spectrum. An exact implementation of this  is given by the
following equation:

\begin{equation} H T = T
H'                                                   \label{3}
\end{equation}

\noindent which imply that if $\Psi '_{E'}$ is an eigenstate of $H'$
then  $T\Psi '_{E'}$ is an eigenstate of $H$ with the same eigenvalue
(unless it belongs to the kernel of $T$: $T\Psi '_{E'}=0$), indeed,

\begin{equation} HT\Psi '_{E'} =  T H'\Psi '_{E'} = E' T\Psi
'_{E'}                                                  \label{4}
\end{equation}

To avoid the possibility that $T\Psi '=0$ with $\Psi '\neq 0$, we
shall impose on $T$ the condition,

\begin{equation} T^{\dagger } T = 1_{\cal
H'}                                                   \label{5}
\end{equation}

\noindent such that

\begin{equation} \Psi = T\Psi '  \Rightarrow \Psi ' = T^{\dag}
\Psi                                                  \label{6}
\end{equation}

\noindent Condition (\ref{5}) thus stablishes a one to one relation
between $\cal H'$ and Im($T$)  in $\cal H$.

\noindent Observe that Eq. (\ref{3}) is nothing but the commutativity
of the following  diagram:

\begin{center} \[ \begin{array}{llcll}
  & {\cal H'} & \stackrel{T}{\longrightarrow} & {\cal H}  &  \\
 H' & \downarrow &     &  \downarrow & H  \\
  &  {\cal H'}  &  \stackrel{T}{\longrightarrow} & {\cal H} &
\end{array} \] \end{center}

 Eqs. (\ref{3}) and (\ref{5}) characterize what may be called exact
renormalization group method  (ERG) in the sense that  the whole
spectrum of $H'$ is mapped onto a part (usually the bottom part)  of
the spectrum of $H$. In practical cases though the exact solution of
Eqs. (\ref{3}) and (\ref{5})  is not possible so that one has to
resort to approximations (see later on). Considering  Eqs. (\ref{3})
and (\ref{5}) we can set up the effective Hamiltonian $H'$ as:

 \begin{equation} H' = T^{\dag} H
T                                                   \label{10}
\end{equation}

\noindent This equation does not imply that the eigenvectors of  $H'$
are mapped onto eigenvectors  of $H$. Notice that Eq.(\ref{10})
together with (\ref{5}) does not imply Eq. (\ref{3}). This happens
because the converse of Eq.(\ref{5}), namely $TT^{\dag } \neq 1_{\cal
H}$  is not true, since otherwise this equation together with
(\ref{5}) would imply that the Hilbert spaces ${\cal H}$ and
${\cal H'}$
are isomorphic while on the other  hand the truncation inherent to the
RG method assumes that $dim {\cal H'} < dim {\cal H}$.

\noindent The fact that $T^{\dag } T\neq 1_{{\cal H}}$ reflects
nothing but the irreversibility of the RG-transformation. Indeed,
we go from ${\cal H}$ to ${\cal H}'$ as prescribed by eq. (\ref{10})
but we cannot reverse that equation.

What Eq.(\ref{10}) really implies is that the mean energy of $H'$ for
the states $\Psi '$ of $\cal H'$  coincides with the mean energy of
$H$ for those states of $\cal H$ obtained through the embedding  $T$,
namely,

  \begin{equation} <\Psi '|H'|\Psi'> =  <T\Psi ' |H| T\Psi'>
\label{11} \end{equation}

In other words $ T\Psi'$ is used as a variational state for the
eigenstates of the Hamiltonian $H$. In  particular $T$ should be
chosen in such a way  that the states truncated in $\cal H$ , which go
down  to $\cal H'$, are the ones expected to contribute the most to
the ground state of $H$. Thus Eq.  (\ref{10}) is the basis of the so
called variational renormalization group method (VRG).
 As a matter  of fact, the VRG method was the first one to be
proposed.  The ERG came afterwards as a perturbative extension of the
former (see later on).

\noindent  More generally, any operator $\cal O$ acting in $\cal H$
can be ``pushed down" or renormalized to a new  operator $\cal O'$
which acts in $\cal H'$ defined by the formula,

  \begin{equation} {\cal O}'  = T^{\dag} {\cal O}
T                                       \label{43} \end{equation}

\noindent Notice that Eq.(\ref{10}) is a particular case of this
equation if choose {\cal O} to  be the Hamiltonian $H$.

In so far we have not made use of the all important concept of the
block,  but a practical implementation of the VRG or ERG methods does
require it. The central role played by this concept  makes all the
real-space RG-methods to be block methods.

Once we have established the main features of the RG-program,  there
is quite freedom to implement specifically these fundamentals. We may
classify this freedom in two aspects:

\begin{itemize}

\item The choice of how to reduce the size of the lattice.

\item The choice of how many states to be retained in the truncation
procedure.

\end{itemize}

\noindent We shall address the first aspect now. There are mainly two
procedures to reduce the  size of the lattice:

\begin{itemize}

\item by dividing the lattice into blocks with $n_s$ sites each. This
is the blocking method  introduced by Kadanoff to treat spin lattice
systems.

\item by retrieving site by site of the lattice at each step of the
RG-program. This is the procedure used by Wilson in his RG-treatment
of the Kondo problem. This method is clearly more suitable  when the
lattice is one-dimensional.

\end{itemize}

  We shall be dealing with the Kadanoff
 block methods mainly because they are well suited to perform
analytical computations and because they are conceptually easy to be
extended to higher  dimensions. On the contrary, the DMRG method
introduced by White \cite{white}  works with the Wilsonian numerical
RG-procedure what makes it intrinsically one-dimensional  and
difficult to be generalized to more dimensions.

The first step of the BRG method consists in asembling the set of
lattice points into diconnected blocks of $n_s$ sites each.

In this fashion there are a total of $N'=N/n_s$ blocks in the whole
chain. This partition of the lattice into blocks induces a
decomposition of the Hamiltonian (\ref{1})  into an intrablock
Hamiltonian  $H_B$ and a interblock Hamiltonian $H_{BB}$:

 \begin{equation}
 H = H_B + \lambda H_{BB}                    \label{21}
 \end{equation}

\noindent where $\lambda $ is a coupling constant which is already
present in $H$ or else it can  be introduced as a parameter
characterizing the interblock coupling and in this latter case one
can  set it to one at the end of the discussion.

\noindent Observe that the block Hamiltonian $H_B$  is a sum of
commuting Hamiltonians each  acting on every block. The
diagonalization of $H_s$ can thus be achieved for small $n_s$ either
analytically or numerically.

\noindent Eq. (\ref{21}) suggests that we should search for solutions
of the intertwiner equation  (\ref{3}) in the form of a perturbative
expansion in the interblock coupling constant parameter  $\lambda $,
namely,

 \begin{equation} T = T_0 + \lambda T_1 +  \lambda^2  T_2 +
\ldots                    \label{22a} \end{equation}

 \begin{equation} H' =H'_0 + \lambda H'_1 +  \lambda^2 H'_2 +
\ldots                    \label{22b} \end{equation}

\noindent To zeroth order in $\lambda $ Eq. (\ref{3}) becomes

 \begin{equation} H_B T_0 = T_0
H_0'                                                 \label{23}
\end{equation}

\noindent Since $H_B$ is a sum of disconnected block Hamiltonians
$h_{j'}^{(B)}$, $j'=1,\ldots ,N'$ implicitly defined through the
relation

 \begin{equation} H_B = \sum
_{j'=1}^{N'}h_{j'}^{(B)}
\label{24} \end{equation}

\noindent one can search for a solution of $T_0$ in a factorized form

 \begin{equation} T_0 = \prod _{j'=1}^{N'}
T_{0,j'}                                                \label{25}
\end{equation}

\noindent and an effective Hamiltonian $H_0'$ which acts only at the
site $j'$ of the new chain,

 \begin{equation} H_0' =   \sum _{j'=1}^{N'}h_{j'}^{(s')} =
H'_{s'}                                                \label{26}
\end{equation}

\noindent Observe that $ H'_{s'}$ is nothing but a site-Hamiltonian
for the new chain. Eq. (\ref{23}) becomes for each block:

  \begin{equation} h_{j'}^{(B)}  T_{0,j'}   =
T_{0,j'}h_{j'}^{(s')}
\label{27} \end{equation}

\noindent The diagonalization of $h_{j'}^{(B)}  $ for $j'=1,\ldots
,N'$ will allow us to write

  \begin{equation} h_{j'}^{(B)} =  \sum _{i=1}^{k'} |i\rangle_{j'} \
\epsilon _i \ _{j'} \hspace*{-1pt} \langle i| +  \sum _{\alpha
=1}^{k^{n_s}-k'} |\alpha \rangle_{j'}  \ \epsilon _{\alpha }  \ _{j'}
\hspace*{-1pt} \langle \alpha |  \label{28} \end{equation}

\noindent where $|i>_{j'}$ for $j=1,\ldots ,k'$ are the $k'$-lowest
energy states of $h_{j'}^{(B)}$. Moreover, we suppose that
$h_{j'}^{(B)}$ is the same Hamiltonian for each block so that
$\epsilon _i$ does not depend on the block.

\noindent The truncated Hamiltonian $h_{j'}^{(s)}$ and the intertwiner
operator $T_{0,j'}$ are then  given by:

  \begin{equation} h_{j'}^{(s')} =  \sum _{i=1}^{k'} |i\rangle'_{j'} \
\epsilon _i \ _{j'} \hspace*{-1pt} \langle i|'   \label{29}
\end{equation}

  \begin{equation} T_{0,j'} =  \sum _{i=1}^{k'} |i\rangle'_{j'} \
_{j'} \hspace*{-1pt} \langle i|'       \label{30} \end{equation}

\noindent  Later on we shall show examples of these relations.

 The final outcome of this analysis is that the effective Hamiltonian
$H'$ has a similar  structure to the one we started with,
namely $H$. The
operators involved in $H'_{s'}$ and  $H'_{s's'}$ may by all means
differ from those of $H_{s}$ and $H_{ss}$, but in some cases  the only
difference shows up as a change in the coupling constants. This is
known as {\em the renormalization of the bare coupling constants}. When
this is the case, one may easily iterate  the RG-transformation and
study the RG-flows.

\Section{Block RG-Approach to the Heisenberg-Ising Model }

To exemplify the standard BRG-method we shall study a 1d-lattice
Hamiltonian,  the Heisenberg-Ising model.
Prior to considering in detail the Antiferromagnetic
(AF) Heisenberg model we shall make
some general considerations concerning Hamiltonians which commute with
a symmetry group ${\cal G}$. Notice that for the  AF Heisenberg model
${\cal G}$ is nothing but the rotation group $SU(2)$. Let us call $g$
an element of the group ${\cal G}$ and $\Pi _{\cal H} (g)$ a
representation of $g$ acting on the Hilbert space  ${\cal H}$. We say
that ${\cal G}$ is a symmetry group of the Hamiltonian $H$ if

\begin{equation} [H ,  \Pi _{\cal H} (g) ] = 0 \ \ \ \ \  \forall g
\in    {\cal G}             \label{11xh1} \end{equation}

\noindent Similarly we want the effective Hamiltonian $H'$ to be
invariant under the action of  ${\cal G}$ acting now on the Hilbert
space ${\cal H'}$,

\begin{equation} [H' ,  \Pi _{\cal H'} (g) ] = 0 \ \ \ \ \  \forall g
\in    {\cal G}             \label{11xh2} \end{equation}

\noindent For this to be the case, the RG-transformation must preserve
the symmetry of the original  hamiltonian. This can be simply achieved
if one choses $T$ as the intertwiner or Clebchs-Gordan  operator.
Indeed we may recall that if a representation say $\Pi _{V_3}$ is
contained in the  tensor product $\Pi _{V_1} \otimes \Pi _{V_2}$ then
one can define the CG-operator as,

\begin{equation} C^{12}_3: \ \ \ \ V_3 \longrightarrow V_1 \otimes
V_2                               \label{11xh3} \end{equation}

\noindent which in fact satisfies the intertwiner condition:

\begin{equation} (\Pi _{V_1} \otimes \Pi_{V_2}) (g)   \ C^{12}_3 =
C^{12}_3 \Pi _{V_3} (g)          \label{11xh4} \end{equation}

\noindent This equation expresses the commutativity of the following
diagram,

\begin{equation} \begin{array}{llcll}
 & V_3    &  \stackrel{C^{12}_3}{\longrightarrow}  &  V_1\otimes V_2
&  \\ \Pi _3 & \downarrow  &   &   \downarrow  &  \Pi _1 \otimes \Pi
_2  \\      \label{11xh5}
 &  V_3  & \stackrel{C^{12}_3}{\longrightarrow}    &  V_1\otimes V_2
&   \end{array} \end{equation}

\noindent Thus from a theoretical point of view, {\em the truncation
process  pertaining to the real-space  RG is nothing but a tensor
product decomposition of representations of the group} ${\cal G}$. To
make this point more explicit let us suppose that $\cal H$ and ${\cal
H'}$ are given as tensor  products as:

\begin{equation} {\cal H}  = \otimes _1^N   V      \label{11xh6a}
\end{equation}

\begin{equation} {\cal H'}  = \otimes _1^{N'}   V'     \label{11xh6b}
\end{equation}

\noindent where $V$ and $V'$ are irreducible representation spaces of
${\cal G}$. Then the block  method of the previous section
 applied to this case
is equivalent to the tensor product decomposition:

\begin{equation} V \otimes \stackrel{n_s}{\ldots } \otimes V
\longrightarrow V'      \label{11xh7} \end{equation}

In Eq. (\ref{11xh7})  one is establishing that the irrep $\Pi _{V'}$
is contained in the tensor product of $N$ copies
 of the irrep $\Pi _V$.

Obviously, the tensor product decomposition usually
contains different irreps. The
criterion to choose a  particular irrep, or a collection of irreps is
the one of {\em minimum energy}. All the states of a  given irrep $V'$
will have the same energy.

The summary of the discussion so far is that {\em
the intertwiner operator
$T_0$ can be identified with  the Clebsch-Gordan operator},

\begin{equation} T_0 = C^{V \otimes \stackrel{n_s}{\ldots } \otimes
V}_{V'}  :  V' \longrightarrow   V \otimes \stackrel{n_s}{\ldots }
\otimes V  \label{11xh8} \end{equation}

Let us illustrate this ideas with the AF Heisenberg-Ising model whose
Hamiltonian is given by:

\begin{equation} H_N = J \sum _{j=1}^{N-1} (S^x_j S^x_{j+1} + S^y_j
S^y_{j+1} + \Delta S^z_j S^z_{j+1})   \label{11xh9} \end{equation}

\noindent where $\Delta \geq 0$ is the anisotropic parameter and $J>0$
for the antiferromagnetic  case. If $\Delta = 1$ one has the
AF-Heisenberg model which was solved by Bethe in 1931. If  $\Delta =
0$ one has the XX-model which can be trivially solved using a
Jordan-Wigner transformation which maps it onto a free fermion model.
For the remaining values of $\Delta $ the model is also  solvable by
Bethe ansatz and it is the 1D relative of the 2D statistical
mechanical model known as
 the 6-vertex or XXZ-model.

\noindent The region $\Delta > 1$ is massive with a doubly degenerate
ground state  in the thermodynamic limit $N \rightarrow \infty$
characterized  by the non-zero value of the staggered magnetization,

\begin{equation} m_{\mbox{st}} = \langle \frac{1}{N} \sum_j S^z_j
(-1)^j  \rangle   \label{11xh9b} \end{equation}

\noindent The region $0 \leq \Delta \leq 1$ is massless and the ground
state is non-degenerate  with a zero staggered magnetization. The
phase transition between the two phases has an  essential singularity.

We would like next to show  which of these features are captured by a
real-space RG-analysis. The rule of thumb for the RG-approach to
half-integer spin model or fermion model is to consider  blocks with
an {\em odd number of sites}. This allows in principle, although not
necesarilly, to  obtain effective Hamiltonians with the same form as
the original ones. Choosing for (\ref{11xh9}) blocks  of 3 sites we
obtain the block Hamiltonian:

\[ \frac{1}{J} H = \vec{S}_1 \cdot \vec{S}_2 +   \vec{S}_2 \cdot
\vec{S}_3 +  \epsilon (S^z_1 S^z_2 + S^z_2 S^z_3)   \] \begin{equation}
= \frac{1}{2} \left\{ [\vec{S}_1 + \vec{S}_2 +\vec{S}_3]^2 -
 (\vec{S}_1 + \vec{S}_3)^2 - 3/4 \right\}   + \epsilon (S^z_1 S^z_2 +
S^z_2 S^z_3)  \label{11xh10} \end{equation}

\noindent $\epsilon \equiv \Delta - 1$.

If $\epsilon = 0$ the block Hamiltonian $H_B$ is invariant under the
$SU(2)$ group and according  to the introduction to this section,  we
should consider the tensor product decomposition:

\begin{equation} \frac{1}{2} \otimes \frac{1}{2} \otimes \frac{1}{2}
= \frac{1}{2}  \oplus \frac{1}{2}  \oplus
\frac{3}{2}
\label{11xh11} \end{equation}

\noindent The particular way of writing $H_B$ given in Eq.
(\ref{11xh10}) suggests to compose first  $\vec{S}_1$ and  $\vec{S}_3$
and then, the resulting spin with $\vec{S}_2$. The result   of
this compositions is given as follows:

\begin{equation} | \frac{3}{2},  \frac{3}{2} \rangle = | \uparrow
\uparrow \uparrow \rangle,  \ \ E_B = J/2       \label{11xh12a}
\end{equation}

\begin{equation} | \frac{3}{2},  \frac{1}{2} \rangle =
\frac{1}{\sqrt{3}}  ( | \uparrow \downarrow \uparrow \rangle  +  |
\downarrow \uparrow \uparrow \rangle  +
   | \uparrow \uparrow \downarrow \rangle  ),     \ \ E_B =
J/2             \label{11xh12b} \end{equation}

\begin{equation} | \frac{1}{2},  \frac{1}{2} \rangle_1 =
\frac{1}{\sqrt{2}}
 (| \uparrow \uparrow \uparrow \rangle - | \downarrow \uparrow
\uparrow \rangle),  \ \ \ E_B =
0
\label{11xh12c} \end{equation}

\begin{equation} | \frac{1}{2},  \frac{1}{2} \rangle_0 =
\frac{1}{\sqrt{6}}  ( 2 | \uparrow \downarrow \uparrow \rangle  -  |
\downarrow \uparrow \uparrow \rangle  -
   | \uparrow \uparrow \downarrow \rangle  ),     \ \ E_B =
-J             \label{11xh12d} \end{equation}

\noindent Hence for $\epsilon =0$
we could choose the spin $1/2$ irrep.
with basis vectors  $|\frac{1}{2}, \frac{1}{2} \rangle_0$ and
$|\frac{1}{2}, -\frac{1}{2} \rangle_0$ in order to define the
intertwiner operator $T_0$.

\noindent However, if $\epsilon \neq 0$ the states (\ref{11xh12a})
-(\ref{11xh12d}) are not eigenstates of  (\ref{11xh10}). The full
rotation group is broken down to the rotation around the z-axis. The
states
  $| \frac{3}{2},  \frac{1}{2} \rangle$ and $| \frac{1}{2},
\frac{1}{2} \rangle_0 $ are mixed in the new ground state which is
given by:

\begin{equation} | + \frac{1}{2} \rangle = \frac{1}{\sqrt{1+2 x^2}}  (
2 | \frac{1}{2},  \frac{1}{2} \rangle_1 + \sqrt{2} x| \frac{3}{2},
\frac{1}{2} \rangle ) \label{11xh13a} \end{equation}

\noindent where

\begin{equation} x = \frac{ 2 (\Delta -1)}{8 + \Delta + 3
\sqrt{\Delta^2 + 8}}              \label{11xh13b} \end{equation}

\noindent and its energy is,

\begin{equation} E_B = -\frac{J}{4} [\Delta + \sqrt{\Delta ^2 +
8}]                                       \label{11xh13c}
\end{equation}

\noindent along with its $| - \frac{1}{2} \rangle$ partner. This are
now the two states retained in the  RG method. To be more explicit, we
have

\begin{equation} | + \frac{1}{2} \rangle = \frac{1}{\sqrt{6(1+2 x^2)}}
[(2 x + 2) |\uparrow \downarrow \uparrow \rangle +  (2 x - 1)
|\uparrow \uparrow \downarrow \rangle +  (2 x - 1) |\downarrow
\uparrow \uparrow \rangle ] \label{11xh14a} \end{equation}

\begin{equation} | - \frac{1}{2} \rangle = -\frac{1}{\sqrt{6(1+2
x^2)}}  [(2 x + 2) |\downarrow \uparrow \downarrow \rangle +  (2 x -
1) |\downarrow \downarrow \uparrow \rangle +  (2 x - 1) |\uparrow
\downarrow \downarrow \rangle ]
\label{11xh14b} \end{equation}

\noindent The intertwiner operator $T_0$ reads then,

\begin{equation} T_0 = | + \frac{1}{2}\rangle \langle  \uparrow |'
+     | - \frac{1}{2}\rangle \langle  \downarrow
|'                                   \label{11xh15} \end{equation}

\noindent where $|\uparrow \rangle'$  and $|\downarrow \rangle'$ form
a basis for the space  $V'={\cal C}^2$. The RG-equations for the spin
operators $\vec{S}_i$ ($i=1,3$) are then given by

\begin{equation} T_0^{\dag} \vec{S}^x_i T_0 =    \xi^x \vec{S' }^x_i
\ \ i = 1,3.                   \label{11xhh15a} \end{equation}

\begin{equation} T_0^{\dag} \vec{S}^y_i T_0 =    \xi^y  \vec{S'
}^y_i     \ \ i = 1,3.                   \label{11xhh15b}
\end{equation}

\begin{equation} T_0^{\dag} \vec{S}^z_i T_0 =    \xi^z  \vec{S'
}^z_i     \ \ i = 1,3.                   \label{11xhh15c}
\end{equation}

\noindent where $\xi^x$, etc are the renormalization factors which
depend upon the anisotropy  parameter by,

\begin{equation} \xi^x = \xi^y
\equiv \frac{2 (1 + x) (1 -2 x)}{3 (1 + 2
x^2)}          \label{11xhh15d} \end{equation}

 \begin{equation} \xi^z
\equiv \frac{2 (1 + x)^2}{3 (1 + 2 x^2)}
\label{11xhh15e} \end{equation}

\noindent Observe the symmetry between the sites $i=1$ and $3$ which
is a consequence of the even parity of the states (\ref{11xh14a})
-(\ref{11xh14b}).

The renormalized Hamiltonian can be easily obtained using
Eqs.(\ref{11xhh15a})-(\ref{11xhh15e})
 and (\ref{11xh9}), and  apart from and additive constant it has the
same form as $H$, namely \cite{rabin},

\begin{equation} T_0^{\dag} H_N (J,\Delta )T_0 =    \frac{N}{3} e_B
(J,\Delta ) +
         H_{N/3} (J',\Delta
')
\label{11xh16} \end{equation}

\noindent where

 \begin{equation} J' = (\xi ^x)^2
J
\label{11xh17a} \end{equation}

 \begin{equation} \Delta' = (\frac{\xi ^z}{\xi^x})^2
\Delta
\label{11xh17b} \end{equation}

\noindent Iterating these equations we generate a family of
Hamiltonians  $H^{(m)}_{N/3^m} (J^(m),\Delta ^{(m)})$. The energy
density of the ground state of $H_N$ in the  limit $N\rightarrow
\infty $ is then given by,

 \begin{equation} \lim _{N\rightarrow \infty} \frac{E_0}{N} =
e^{BRG}_{\infty} =  \sum _{m=0}^{\infty} \frac{1}{3^{m+1}} e_B
(J^{(m)},\Delta^{(m)})          \label{11xh18} \end{equation}

\noindent where initially $J^{(0)}=J$, $\Delta^{(0)}=\Delta$ and
Eqs.(\ref{11xh17a}) -(\ref{11xh17b})  provide the flow of the coupling
constants.

The analysis of Eq.(\ref{11xh17b}) shows that there are 3 fixed points
corresponding to the values  $\Delta =0$ (isotropic XX-model), $\Delta
=1$ (isotropic Heisenberg model) and $\Delta =\infty$  (Ising model).

\noindent The computation of $e^{BRG}_{\infty}$ in this case is
facilitated by the fact that (\ref{11xh18})  becomes a geometric
series at the fixed point. The exact results concerning the models
$\Delta=0$  and $\Delta =1$ are extracted from references
\cite{lieb-s-m} and \cite{orbach}. The case with  $\Delta \rightarrow
\infty$ is exact because the states $|\pm \frac{1}{2} \rangle$ given
in  (\ref{11xh14a}) - (\ref{11xh14b}) tend in that limit to the exact
ground state $|\uparrow \downarrow \uparrow \rangle$ and  $|\downarrow
\uparrow \downarrow \rangle$ of the Ising model. As a matter of fact,

\[  |+\frac{1}{2} \rangle \simeq _{\Delta \rightarrow \infty}
|\uparrow \downarrow \uparrow \rangle -  \frac{1}{\Delta} |\uparrow
\uparrow \downarrow \rangle -  \frac{1}{\Delta} |\downarrow \uparrow
\uparrow \rangle \]

 \[  |-\frac{1}{2} \rangle \simeq _{\Delta \rightarrow \infty}
-|\downarrow \uparrow \downarrow \rangle -  \frac{1}{\Delta}
|\downarrow \downarrow \uparrow \rangle -  \frac{1}{\Delta} |\uparrow
\downarrow \downarrow \rangle \]

The region $0 < \Delta < 1 $ which flows under the RG-transformation
to the XX-model is massless  since both $J^{(m)}$ and $\Delta ^{(m)}$
go to zero. We showed at the begining of this section  that all this
region is critical (a line of fixed points) and therefore massless.
The RG-equations  (\ref{11xh17a}) -(\ref{11xh17b}) are not able to
detect this criticality except at the point $\Delta =0$. Only  the
masslessness property is detected.

The region $\Delta >1$ which flows to the Ising model is massive and
this follows from the fact  that the product $J^{(m)} \Delta ^{(m)}$
goes in the limit $m \rightarrow \infty$ to a constant quantity
$J^{(\infty)} \Delta ^{(\infty)}$ which can be computed from Eqs.
(\ref{11xh17a}) -(\ref{11xh17b}) and  (\ref{11xh15}),

 \begin{equation}
  J^{(\infty)} \Delta ^{(\infty)} = \prod _{m=0}^{\infty}
\frac{4}{9}  \frac{(1 + x_m)^4}{(1 + 2 x_m^2)^2}   \label{11xh19}
\end{equation}

\noindent where $x_m$ is given by (\ref{11xh13b}) with $\Delta $
replaced $\Delta ^{(m)}$. This quantity  gives essentially the mass
gap above the ground state and also the end-to-end or LRO order (Long
Range Order) given by the expectation value $|\langle \vec{S}(1) \cdot
\vec{S}(N) \rangle|$ in the limit  $N \rightarrow \infty$.

In summary, the properties of the Heisenberg-Ising model are
qualitatively and quantitatively well  described in the massive region
$\Delta >1$ while in the massless region $0 < \Delta < 1$ one
predicts the massless spectrum {\em but no criticality at each value
of $\Delta $}. This latter  fact is rather subtle and elusive. One
would like to construct a RG-formalism such that the  Hamiltonian
$H_N(\Delta )$ would be a fixed point Hamiltonian for every value of
$\Delta $ in the  range from -1 to 1.
 Hence we postpone this  discussion to next section.

  The phase transition between the two regimes is correctly predicted
to happen  at the value  $\Delta = 1$. This is a consequence of the
rotational symmetry, namely at $\Delta = 1$ the  system is $SU(2)$
invariant and the RG transformation has been defined as to preserve
this  symmetry. When $\Delta \neq 1$ the $SU(2)$ symmetry is broken
and this is reflected later on  in the RG-flow of the coupling
constant $\Delta $.
 The region $0 < \Delta < 1$ corresponds to a central charge
$c=1$, namely, it is realize by a boson compactified in a circle
which radius depends on $\Delta $.
 We may wonder whether the criticality of the region $|\Delta | \leq
1$ is due to some non-trivial  symmetry underlying the anisotropic
Hamiltonian $H$.

\Section{Quantum Groups and the Block Renormalization Group Method
for the Heisenberg-Ising Model}

We present in this section a novel treatment of the Block
Renormalization Group method  for one dimensional quantum Hamiltonians
based on the introduction of a {\em
quantum group}. Our aim is to address the important questions left
open in the  previous section and to clarify the peculiar role played
by the Renormalization group in the  one dimensional physics
\cite{q-germanyo}.

Let us consider the following open spin chain Hamiltonian,

 \begin{equation} H_N(q,J) = \frac{J}{4} \left\{ \sum _{j=1}^{N-1}
\sigma^x_j \sigma^x_{j+1} +  \sigma^y_j \sigma^y_{j+1}  + \frac{q +
q^{-1}}{2} \sigma^z_j \sigma^z_{j+1}  - \frac{q - q^{-1}}{2} (
\sigma^z_1  - \sigma^z_{N} )
\right\}
\label{11xh20} \end{equation}

\noindent where $q$ is an arbitrary quantum parameter.
This Hamiltonian is known to be integrable
\cite{alcaraz}, \cite{gaudin}, \cite{sklyanin}, \cite{cherednik}.
In an interesting paper Pasquier and Saleur \cite{pasquier-saleur}
established the $q$-group invariance of (\ref{11xh20}) which
served to get a better understanding of the interplay between
$q$-groups and CFT at a discrete or lattice level. We have already
talked about the relation between
$q$-groups and CFT in the introduction when reference \cite{ags}
was mentioned, but this concerned the models in the continuum.
What we would like to show now is that the real-space RG method
applied to (\ref{11xh20}) may be perhaps the way to link both
the discrete and continuous approaches between the relation of
$q$-groups and CFT.

\noindent
 As a
matter of fact,
(\ref{11xh20}) is invariant under the following quantum group
generators \cite{pasquier-saleur}: $S^+, S^-$ and $S^z$,

 \begin{equation} S^z = \frac{1}{2} \sum _{j=1}^{N}
\sigma^z_j                                          \label{11xh21a}
\end{equation}

 \begin{equation} S^{\pm } = \sum _{j=1}^{N} q^{-\frac{1}{2} (\sigma
_1^z + \cdots \sigma _{j-1}^z)}
 \sigma^{\pm }_j  q^{\frac{1}{2} (\sigma _{j+1}^z + \cdots \sigma
_{N}^z)}             \label{11xh21b} \end{equation}

\noindent which satisfies the quantum group algebra:

 \begin{equation} [S^+, S^-] = \frac{q^{2 S^z} - q^{-2 S^z}}{q -
q^{-1}}      \label{11xh22a} \end{equation}

 \begin{equation} [S^z, S^{\pm}] = \pm S^{\pm}     \label{11xh22b}
\end{equation}

\noindent In the limit $q \rightarrow 1$ one recovers from
Eqs.(\ref{11xh21a}) -(\ref{11xh22b}) the usual  algebra and addition
rules of $su(2)$. For $q\neq1$ this algebra is  the {\em quantum
universal
 enveloping algebra} ${\cal U}_q (su(2))$, or simply the quantum
$su(2)$ group denoted by $SU_q(2)$. The important property is that the
generators (\ref{11xh21a})-(\ref{11xh21b}) commute with the
Hamiltonian  (\ref{11xh20}):

  \begin{equation} [H_N(q), S^{\pm}] = [H_N(q), S^{z}]     =
0                     \label{11xh23a} \end{equation}

\noindent If we set

   \begin{equation} \Delta = \frac{q + q^{-1}}{2}
\label{11xh23} \end{equation}

\noindent we observe that the bulk terms of Eq.(\ref{11xh20})  and the
one in Eq.(\ref{11xh9}) coincide. The only difference appears in the
boundary term of Eq.(\ref{11xh20}) which is essential for the
existence of the  quantum  group symmetry.

\noindent There are two important cases which we can consider:

   \begin{equation} q: \mbox{real and positive }  \ \ \Rightarrow \ \
\Delta \geq 1           \label{11xh28a} \end{equation}

    \begin{equation} |q| = 1  \ \ \Rightarrow \ \  |\Delta | \leq
1           \label{11xh28b} \end{equation}

\noindent If $q$ is real then $H_N(q)$ is Hermitean while if $q$ is a
phase then $H_N(q)$ is not  Hermitean but nevertheless its spectrum is
real for $H_N(q)$ and $H_N(q^{-1})$ are related by a  similarity
transformation. This case is the most interesting one. In particular
if we write $q$ as  a root of unity $q=e^{\frac{i\pi}{\mu + 1}}$, then
the Hamiltonian (\ref{11xh20}) is a critical Hamiltonian  with a
Virasoro central algebra given by,

     \begin{equation} q = e^{\frac{i\pi}{\mu + 1}}  \ \ \Rightarrow \
\  c = 1 - \frac{6}{\mu (\mu + 1)}           \label{11xh29}
\end{equation}

\noindent However for the time being let us keep $q$ as an arbitrary
parameter characterizing the  anisotropy of the model (\ref{11xh23}).

Let us try to apply the Block Renormalization Group Method to the
analysis of the Hamiltonian  (\ref{11xh20}). To this end we have to
write (\ref{11xh20})  in the following form:

      \begin{equation} H_N(q) = \sum _{j=1}^{N-1} h_{j,j+1}
(q,J)                                                 \label{11xh24a}
\end{equation}

      \begin{equation}
 h_{j,j+1} (q,J) =  \frac{J}{4} [\sum _{j=1}^{N-1} \sigma^x_j
\sigma^x_{j+1} +  \sigma^y_j \sigma^y_{j+1}  + \frac{q + q^{-1}}{2}
\sigma^z_j \sigma^z_{j+1}  - \frac{q - q^{-1}}{2} ( \sigma^z_j  -
\sigma^z_{j+1} ) ]                             \label{11xh24b}
\end{equation}

\noindent Observe that each boundary term in $h_{j,j+1}$ cancels one
another leaving only those  at the end of open chain (i.e. $j=1$ and
$N$). The nice feature of the site-site Hamiltonians  (\ref{11xh24b})
is that all of them commute independently with the $q$-group
generators $S^{\pm }$ and  $S^z$;

  \begin{equation} [h_{j,j+1}, S^{\pm}] = [h_{j,j+1}, S^{z}] = 0\ \ \
\ \forall j = 1,\ldots ,N-1    \label{11xh25} \end{equation}

\noindent Using $h_{j,j+1}$ we can construct $q$-group invariant block
Hamiltonians. If the block  has for example 3 sites then we will have,

  \begin{equation} H_B = h_{1,2}  +
h_{2,3}
\label{11xh26} \end{equation}

\noindent In the isotropic case (i.e. $q=1$) we employ the
Clebsch-Gordan decomposition (\ref{11xh11})  in order to find the
eigenstates of the block Hamiltonian $H_B$ ($\epsilon = 0$) given in
(\ref{11xh10}). For quantum groups, we can also perform $q$-CG
decompositions. The new feature is  that now the $q$-CG coefficients
depend on the value of $q$. For generic values of $q$  the analoge of
Eqs.(\ref{11xh12a}) -(\ref{11xh12d}) are given by:

\begin{equation} | \frac{3}{2},  \frac{3}{2} \rangle = | \uparrow
\uparrow \uparrow \rangle
\label{11xh27a} \end{equation}

\[  e_B = J \frac{q + q^{-1}}{4}        \]

\begin{equation} | \frac{3}{2},  \frac{1}{2} \rangle =
\frac{1}{\sqrt{[3]_q}}  (  | \uparrow \downarrow \uparrow \rangle  +
q | \downarrow \uparrow \uparrow \rangle  +
   q^{-1}| \uparrow \uparrow \downarrow \rangle  )
                                            \label{11xh27b}
\end{equation}

\[ e_B = J \frac{q + q^{-1}}{4}                    \]

\begin{equation} | \frac{1}{2},  \frac{1}{2} \rangle_1 =
\frac{1}{\sqrt{2 (q + q^{-1} - 1)}}
 ( q^{1/2} | \uparrow \uparrow \downarrow \rangle  + (q^{1/2} -
q^{-1/2}) |\uparrow \downarrow \uparrow \rangle - q^{-1/2} |
\downarrow \uparrow \uparrow \rangle)
\label{11xh27c} \end{equation}

\[
 e_B = J \frac{2 - q - q^{-1}}{4}           \]

\begin{equation} | \frac{1}{2},  \frac{1}{2} \rangle_0 =
\frac{1}{\sqrt{2 (q + q^{-1} + 1)}}  ( (q^{1/2} + q^{-1/2}) | \uparrow
\downarrow \uparrow \rangle  -  q^{-1/2}  | \downarrow \uparrow
\uparrow \rangle  -
  q^{1/2}  | \uparrow \uparrow \downarrow \rangle  )
\label{11xh27d} \end{equation}

\[
 e_B = -J \frac{2 + q + q^{-1}}{4}        \]

\begin{equation} | \frac{1}{2},  -\frac{1}{2} \rangle_0 =
\frac{1}{\sqrt{2 (q + q^{-1} + 1)}}  ( -(q^{1/2} + q^{-1/2}) |
\downarrow \uparrow \downarrow \rangle  -  q^{1/2}  | \downarrow
\downarrow \uparrow \rangle  -
  q^{1/2}  | \downarrow \downarrow \uparrow \rangle  )
\label{11xh27e} \end{equation}

\[
 e_B = -J \frac{2 + q + q^{-1}}{4}          \]

\noindent where $[3]_q$ denotes the $q$-number with the usual
definition

\begin{equation} [n]_q = \frac{q^n - q^{-n}}{q - q^{-1}}    \ \ n \in
{\cal N}                    \label{11xhh27} \end{equation}

The normalization of the states in Eqs. (\ref{11xh27a}) -
(\ref{11xh27e}) is as if $q$ were always a real number, a bra vector
$\langle |$ means transposing a ket vector.

If $q$ goes to 1 the vectors in Eqs. (\ref{11xh27a}) - (\ref{11xh27e})
go over the vectors in  Eqs. (\ref{11xh12a}) - (\ref{11xh12d}). On the
other hand, for $q\neq 1$ there is a certain similarity between  the
states $|\frac{1}{2},\pm \frac{1}{2} \rangle_0$ and the states   $|\pm
\frac{1}{2} \rangle$ given in Eqs. (\ref{11xh14a}) - (\ref{11xh14b}).
However the main difference is  that   $|\frac{1}{2},\pm \frac{1}{2}
\rangle_0$ are not parity invariant  $(1\leftrightarrow 3), (2
\leftrightarrow 2)$, while states (\ref{11xh14a}) - (\ref{11xh14b})
are.

If we let $q$ go to zero then $\Delta$ goes to $+\infty $. Let us
recall that in this limit the states  $|\pm \frac{1}{2} \rangle$
(\ref{11xh14a}) - (\ref{11xh14b}) go to the exact ground state of the
Hamiltoninan  (\ref{11xh10}). However in the case of the states
$|\frac{1}{2},\pm \frac{1}{2} \rangle_0$ one does not  recover the
exact eigenstates in this limit. This shows that the states
(\ref{11xh27a}) - (\ref{11xh27e})  are not appropriate for a discussion
of the AF region $\Delta > 1$ (or $q$ real). Hence we shall  confine
ourselves to the critical region $|\Delta| < 1$ where $q$ is a pure
phase. As we did already  for the isotropic case we shall truncate the
basis (\ref{11xh27a}) - (\ref{11xh27e}) to the states
$|\frac{1}{2},\pm \frac{1}{2} \rangle_0$ and therefore the intertwiner
operator $T_0$ is given by,

\begin{equation} T_0 =   |\frac{1}{2}, \frac{1}{2} \rangle_0 \langle
\uparrow |' +
   |\frac{1}{2}, -\frac{1}{2} \rangle_0 \langle \downarrow |'
\label{11xh30} \end{equation}

\noindent which satisfies the normalization condition,

\begin{equation} T_0^{t } T_0 = 1                 \label{11xh31}
\end{equation}

\noindent where the superscript $t$ stands for the transpose of the
operator (instead of the adjoint).

\noindent This means that we can get the effective Hamiltonian $H'$
through the formula:

\begin{equation} H' = T_0^{t }  H
T_0                                                     \label{11xh32}
\end{equation}

The RG-equations for the spin
operators $\vec{S}_i$ ($i=1,3$) are then given by

\begin{equation} T_0^{t} \vec{S}^x_i T_0 =    \xi  \vec{S' }^x   \ \ i
= 1,3.                   \label{11xhh33a} \end{equation}

\begin{equation} T_0^{t} \vec{S}^y_i T_0 =    \xi  \vec{S' }^y    \ \
i = 1,3.                   \label{11xhh33b} \end{equation}

\begin{equation} T_0^{t} \vec{S}^z_i T_0 =    \xi  \vec{S' }^z  +
\eta_i 1'   \ \ i = 1,3.  \label{11xhh33c} \end{equation}

\noindent where $\xi$ is the renormalization factor which depends upon
the anisotropy  parameter through the $q$-parameter in the following
fashion,

\begin{equation} \xi = \frac{q + q^{-1} + 2}{2 (q + q^{-1} +
1)}          \label{11xhh33d} \end{equation}

\noindent and

 \begin{equation} \eta_1 = - \eta_3  := \eta =  \frac{q - q^{-1} }{4
(q + q^{-1} + 1)}          \label{11xh34} \end{equation}

\noindent Observe that there are quite a few remarkable differences
between this ``quantum"  renormalization prescription Eqs.
(\ref{11xhh33a}) - (\ref{11xhh33d}) with respect to the ordinary
renormalization expressed in Eqs. (\ref{11xhh15a}) -
(\ref{11xhh15e}).  To begin with, the renormalization constant $\xi $
is common to all the spin operators regardless of  its spatial
component.  This is a  reflection of the $SU_q(2)$ preservation of the
RG method adopted. And last but not least, observe  the presence in
Eq.(\ref{11xhh33c}) of an extra term proportional to the identity
operator. We may call  this term a {\em quantum group anomaly}.

Using these equations we can compute the renormalization of the
block-block Hamiltonian $h_{BB}$ which turns out to be of the same
form as the original site-site Hamiltonian (\ref{11xh24b}) with  {\em
the same value of $q$}, namely

 \begin{equation} T^t_0 h_{3 k,3 k + 1} (q,J) T_0 =  \frac{J}{4} \xi^2
[\sigma'^x_k \sigma'^x_{k+1} + \sigma'^y_k \sigma'^y_{k+1} +
 \frac{q + q^{-1}}{2} \sigma'^z_k \sigma'^z_{k+1}
\end{equation}                         \label{11xh35} \[
 - \frac{q - q^{-1}}{2} (\sigma'^z_k  - \sigma'^z_{k+1} ) +  e_{BB}
(q,J)]
\]

\noindent with

 \begin{equation} e_{BB} (q,J) = J \frac{(q - q^{-1})^2 (3q + 3q^{-1}
+ 4)}{32 (q + q^{-1} + 1)^2}       \label{11xh36} \end{equation}

\noindent Combining Eqs.(\ref{11xh35}), (\ref{11xh36})  and
(\ref{11xh27a}) - (\ref{11xh27e}) we finally arrive at quantum group
RG-equations,

 \begin{equation} T^t_0 H_N (q,J) T_0 = H_{N/3} (q',J') + \frac{N}{3}
e_B(q,J) + (\frac{N}{3} - 1)e_{BB} (q,J) \label{11xh37} \end{equation}

 \noindent with

 \begin{equation}
 e_B(q,J) = -J \frac{2 + q +
q^{-1}}{4}
\label{11xh38a} \end{equation}

 \begin{equation} q' =
q
\label{11xh38b} \end{equation}

 \begin{equation} J' = \xi^2
J
\label{11xh38c} \end{equation}

\noindent Hence we obtain a quite remarkable result we were searching
of, namely, that the  coupling constant $\Delta$ or alternatively {\em
q does not flow under the RG-transformation}, while $J^{(m)}$ goes to
zero in the limit when $m \rightarrow \infty$, which in turn implies
that  theory is massless.

The computation of the ground state energy of $H_N$ for any value of
$N$ is very simple since  one has only to compute a geometrical
series. The result is

 \begin{equation} E_0(N) = N \frac{1 - (\xi ^2/3)^M}{3 - \xi ^2} (e_B
+ e_{BB}) -  \frac{1 - \xi^{2M}}{1 - \xi^2}
e_{BB}
\label{11xh39} \end{equation}

 \noindent where $M$ is the number of RG-steps we have to make in
order to resolve the chain of  $N = 3^M$ sites.

\noindent As a check of the validity of this expression and because of
its own interest as well,  we shall consider the case $q = e^{i\pi/3}$
in (\ref{11xh39})  which yields,

 \begin{equation} E_0(N,q=e^{i\pi/3}) =-\frac{3}{8} N +
\frac{3}{8}                       \label{11xh40} \end{equation}

\noindent This expression coincides with the {\em exact} result
obtained through Bethe ansatz  in reference \cite{alcaraz}.
Recall that the finite-size corrections to the free-energy for conformally
invariant
two-dimensional systems behaves as \cite{cardy}, \cite{affleck},

 \begin{equation} E = e N + e_s -
\frac{\pi c}{24}        \frac{1}{N}                 \label{11xh40b}
\end{equation}

\noindent where $e$ is the bulk energy per unit length and $e_s$ is the
surface energy (whihc vanishes for periodic boundary conditions).
 Since the
term proportional to $1/N$ is absent in (\ref{11xh40}) one  observes
that this value of $q$ corresponds to a central extension $c = 0$ of
the Virasoro algebra, in agreement with equation (\ref{11xh29}) ($\mu
= 2$).

At first sight it looks surprising that an approximation method such
as the Block Renormalization  Group yields the exact result at least
in the case $q = e^{i\pi/3}$. The peculiarity of this value of  $q$,
and in general when $q$ is a root of unity, has been noticed in
various contexts  \cite{germancesar}: conformal field theory, quantum
groups and in fact they are intimately related.

\noindent The first thing to be noticed is that at $q = e^{i\pi/3}$
the two denominators of the states  $| \frac{3}{2}, \frac{1}{2}
\rangle$ and $| \frac{1}{2}, \frac{1}{2} \rangle_1$ {\em vanishes}
reflecting the fact that the ``norm" of these states is zero, i.e.,
they are null states and therefore they must be droped out in a
consistent theory. It has been shown in
\cite{pasquier-saleur}
 that because of $(S^{\pm})^3 = 0$ the 6 states  $|\frac{3}{2}, m
\rangle$ and $|\frac{1}{2}, m \rangle_1$ do not form two irreps of
dimensions 4 and  2 but rather a simple indecomposable but not
irreducible representation of $SU_q(2)$. All these  facts motivates
that the tensor product decomposition (\ref{11xh11}) for the case $q =
e^{i\pi/3}$ should  be really be taken as

 \begin{equation} \left( \frac{1}{2} \otimes \frac{1}{2} \otimes
\frac{1}{2} \right)_{q = e^{i\pi/3}} = \frac{1}{2} \label{11xh41}
\end{equation}

\noindent where the irrep 1/2 on the right hand side denotes the one
generated by  $|\frac{1}{2}, m \rangle_0$.

This truncation is mathematically consistent and concides precisely
with the truncation we have  adopted in our Block Renormalization
Group approach to the $q$-group invariant Hamiltonian  (\ref{11xh20}).

{}From a physical point of view, the truncation
(\ref{11xh41})  means that the states
 $| \frac{3}{2}, \frac{1}{2} \rangle$ and $| \frac{1}{2}, \frac{1}{2}
\rangle_1$ are not ``good excited  states" above the ``local ground
state" given by $|\frac{1}{2}, m \rangle_0$. In other words, above the
ground state there are not well behaved excited states. This is why
the  central extension is $c = 0$ which means that the unique state in
the theory is actually the  ground state. What the BRG method does is
to pick up that piece of the ground state which  projects onto a given
block! In the case of $q = e^{i\pi/3}$ we have therefore construct for
chains  with $N = 3^M$ sites the exact ground state of the model
through the BRG method. It is worthwhile to point out that this
derivation is independent of the Bethe ansatz construction  and relies
completely on the quantum group symmetry.

 Another interesting example is
provided by $q=e^{i\pi/4}$ which corresponds to the critical Ising
model ($c=1/2$). The RHS for this $q$ in eq. (\ref{11xh41})
contains two
spin-1/2 irreps. According to the qRG method the truncation of the
spin-3/2 irrep should be a legitimate operation involving {\em no
approximation at all}.
 In references \cite{ags} the
representation
theory of $q$-groups was put in one-to-one
correspondence with that of
Rational
Conformal Field Theories (RCFT). There it was observed that the
truncation
inherent in the construction of the RCFT's has a parallel in the
truncation of
the representation theory of $q$-groups with $q$ a root of unity. The
result we
have obtained in this letter suggests that $q$-group
truncations can be
carried
over a RG analysis of $q$-group invariant chains.
In other words, using
$q$-groups we can safely truncate states in the block RG method.
 We may
summarize
this discussion squematically as we have mentioned in the
introduction.

\Section{qRG Treatment of the ITF Model}

 This simple model has been
widely
used to test the validity of BRG methods \cite{drell}, \cite{jullien}.
The
Hamiltonian of an open chain is given by $H=\sum _{j=1}^{N-1} h_{j,j+1} $
where

\begin{equation} h_{j,j+1} = -(J \sigma_j^x \sigma_{j+1}^x + p \sigma
_j^z + p'
\sigma_{j+1}^z )
\label{16}
\end{equation}

\noindent The standard choice is $p=p'=\Gamma/2$, in which case
(\ref{16}) has 4
different eigenvalues.  The BRG method with a block with two sites
chooses just
the 2 lowest ones. However if $(p,p') = (\Gamma,0)$ (or $(0,\Gamma )$)
the
Hamiltonian (\ref{16}) has two doubly degenerate eigenvalues $\pm e_B$
($e_B=\sqrt{J^2+\Gamma^2}$). This choice is not parity invariant but it
implements the self-duality property of the ITF model, yielding the exact
value
of the critical fixed point of the  ITF which appears at $(\Gamma/J)_c=1$
\cite{fpacheco}. In the following we shall make the choice
$(p,p') = (\Gamma,0)$.
This degeneracy of the spectrum of (\ref{16}) has a {\em
q-group
origin}. The relevant quantum group is again $SU_q(2)$ with $q^4=1$.
However the
representations involved are not a $q$-deformation of the spin $1/2$
irrep. as in
the previous example,  but rather a new class of irreps. which only exist
when
$q$ is a root of unity. They are called {\em cyclic irreps.} and neither
are
highest weight nor lowest weight representations as the more familiar
regular
irreps. If we call $E$, $F$ and $K$ the generators of $SU_q(2)$, which
correspond
essentially to $S^{+}$, $S^-$ and $q^{2 S^z}$ in the notation of the
previous
example, then a cyclic irrep. acting at a single site of the chain is
given by:

\begin{equation} E_j = a \sigma_j^x , \ \ F_j = b \sigma_j^y , \ \ K_j =
\lambda
\sigma_j^z       \label{17} \end{equation}

\noindent where $a=\frac{1}{2} \sqrt{\lambda ^2-1}$, $b=-\frac{1}{2}
\sqrt{1-\lambda ^{-2}}$. The parameter $\lambda $ is the label of the
cyclic
irrep.
Strictly speaking, we have a particular kind of cyclic irreps.
Indeed, the ones which allow the existence of an intertwiner for
their tensor product.
Using (\ref{17}) and the addition rule of $SU_q(2)$ we can get the
representation of $E$, $F$ and $K$ acting on the whole chain:

\begin{eqnarray} E & = & a \sum_{j=1}^N \lambda ^{j-1} \sigma_1^z \cdots
\sigma_{j-1}^z \sigma_j^x   \label{18} \\ F & = & b \sum_{j=1}^N \lambda
^{j-N}
\sigma_j^y \sigma_{j+1}^z \cdots \sigma_{N}^z \\ K & = & \lambda ^N
\prod_{j=1}^N
\sigma_j^z \end{eqnarray}

\noindent Now it is a simple exercise to check that these operators
commute with
(\ref{16}),

\begin{equation} [h_{j,j+1}, E] = [h_{j,j+1}, F] = [h_{j,j+1}, K] = 0, \
\
\forall j                    \label{19} \end{equation}

\noindent assuming that we choose

\begin{equation} \lambda = \Gamma /J                   \label{20}
\end{equation}

\noindent The last of the equalities in (\ref{19}) expresses the
well-known
${\cal Z}_2$-symmetry of the ITF-model which allows one to split the
spectrum of
the Hamiltonian into an even and odd subsectors. {\em The other two
symmetries
are new} and explain the degeneracy of the spectrum of $h_{j,j+1}$. By
all means
the whole Hamiltonian $H=\sum_j h_{j,j+1}$ is also invariant under
(\ref{18}).
Notice that $H$ differs from the standard ITF simply in a term at one of
the ends
of the chain. This is the same mechanism as for the XXZ Hamiltonian: one
needs
properly chosen operators at the boundary in order to achieve quantum
group
invariance. Similarly as for the XXZ model the RG-analysis of the ITF
becomes a
problem in representation of quantum groups: blocking is equivalent to
tensoring
representations. What is the tensor product of cyclic irreps.? Here it is
important to realize that all cyclic irreps. of $SU_q(2)$ have dimension
2, what
distinguishes them is the value of $\lambda $. The tensor product
decomposition
of two cyclic irrep. $\lambda_1$ and $\lambda _2$
of the type given in (\ref{17}) is given by:

\begin{equation} [\lambda _1] \otimes [\lambda _2] = 2 \   [\lambda _1
\lambda
_2]               \label{21b} \end{equation}

\noindent where the 2 means that $\lambda_1 \lambda_2$ appears twice in
the
tensor product. If we perform a blocking of two sites we will get two
cyclic
irreps. corresponding to $\lambda ^2$. Then we expect from $q$-group
representation theory that the new effective Hamiltonian $h'_{j,j+1}$
will have
the same form as (\ref{16}) but with new renormalized coupling constants
$J'$ and
$\Gamma '$ satisfying:

\begin{equation} \lambda' = \frac{\Gamma '}{J'} = (\frac{\Gamma }{J})^2 =
\lambda^2        \label{22} \end{equation}

\noindent This is indeed the result obtained in \cite{fpacheco}. We
arrive
therefore at the conclusion that {\em the RG-flow of the ITF Hamiltonian
(\ref{16}) is equivalent to the tensor product decomposition of cyclic
irreps of
$SU_q(2)$}. This $q$-group interpretation of the RG-flow is independent
of the
size of the blocks: for a $n$-site block the RG-flow would be $\lambda
\rightarrow \lambda^n$. The fixed point $\lambda =1$ of (\ref{22})
describes the
critical regime of the ITF Hamiltonian and it corresponds to a {\em
singular
point in the manifold of cyclic irreps.}\cite{kac-deconcini}.
 At $\lambda =1$ the operators (\ref{18}) are still
symmetries of the Hamiltonian ($a$, $b$ taking any non-zero value) and
they
recall the Jordan-Wigner map between Pauli matrices and 1d-lattice
fermions.

\noindent Of the two equivalent irreps $\lambda^2$ appearing in the
tensor product $\lambda  \otimes \lambda $
we only pick up one of them, which is the lowest energy.
Observe that the loss of information implied by the RG method is
in the ITF case somehow redundant information from the $q$-group
point of view. The RG-flow in  $\lambda $  is in some sense exact
and not affected by the RG-procedure.

\noindent The results we have obtained in the ITF model may perhaps
be realized in more complicated models. Namely, that the space of
coupling constants or equivalently, the space of Hamiltonians of a
given theory
which is where the RG takes place, is the Spec manifold of an
underlying quantum algebra, so that the RG-flow is given by the
tensor product decomposition of the algebra.
For this to work we have to consider quantum groups with a rich
and ``exotic" representation theory, which is indeed the case
when $q$ is a root of unity. In these cases we know from
references \cite{arnaudon}  and \cite{kac-deconcini}
that the Spec of these quantum groups is indeed very rich.
We have used here the simplest situation.
As we said above, it would be interesting to know wheather there are
more complicated realizations of these ideas.
The chiral Potts model is a potential candidate for this realization
due to its well known connection to $SU_q(2)$ with $q^N=1$.

\Section{Conclusions}

In this paper we have presented a brief description of the
Renormalization Quantum Group Method (qRG) which is specially
well-suited to treat 1D quantum lattice Hamiltonians. We have
applied real-space RG methods to study two quantum group invariant
Hamiltonians, namely, the Heisenberg-Ising model and the Ising model
in a transverse field (ITF model). They are defined in an open chain
with appropriate boundary terms. The defining feature of this qRG
method is that the quantum group symmetry is preserved under the RG
transformations except for the appearence of a quantum group anomalous
term which vanishes in the classical case.
We have called it {\em the quantum group anomaly}.

As for the Heisenberg-Ising model, with the aide of
the qRG equations we have shown that the $q$ parameter describing
the anisotropy coupling constant $\Delta$ does not flow under the
RG-transformation when $q$ is of modulus one, while the coupling
constant $J$ goes to zero implying that the theory is massless.
In this fashion, the RG-flow diagram obtained with the qRG method
gives the correct line of critical points exhibited by the exact
model.

In the ITF model, we have shown that the qRG-flow coincides with the
tensor product decomposition of cyclic irreps of $SU_q(2)$ with
$q^4=1$.
Cyclic irreps. were used in \cite{jimbo} to derive the Boltzmann
weights  of the ${\cal Z}_N$-chiral Potts model \cite{macoy}. In
\cite{jimbo} the labels  of the cyclic irreps. have the meaning of
rapidities rather than coupling  constants as in our realization.

The ${\cal Z}_2$ CP-model is nothing but the ITF model. We may wonder
whether the general ${\cal Z}_N$-chiral Potts model admits a $q$-group
RG treatment  along the lines of this work. This problem will be
considered elsewhere. A model  that admits a qRG analysis is the XY
model with a magnetic field $h$ (XYh).  The results will be presented
in \cite{nos}. It suffices to say here that the $q$-group underlying
the model is $SU_q(2)$ with $q^4=1$ and the  representation used are
the so called {\em nilpotent irreps.} \cite{germancesar}, which  are
also described by a parameter $\lambda $ analoge to that in (\ref{17})
and  {\em related to the magnetic field $h$}. The XYh model is
equivalent to a free  fermion with chemical potential. The results we
are obtaining  can be translated  into a $q$-group symmetry between
fermions, either free as in the XY or ITF  models or interacting as
the XXZ model. Another interesting model of interacting  fermions is
the Hubbard model, which has been studied using RG-methods in
\cite{hirsch} . The integrability of the 1D Hubbard model
\cite{liebwu} suggests that it  might be studied using our qRG
techniques.

All the Hamiltonians analysed in this letter are one dimensional, so
the quantum groups are of the type that we know.
 Despite the fact that the Yang-Baxter  equation (the precursor of
$q$-groups) has a higher dimensional analogue called  the
Zamolodchikov or tetrahedron equation \cite{zamo},  the corresponding
high dimensional analogue of quantum groups is not known. This fact
represents  a barrier to a qRG analysis of Hamiltonians defined in
dimensions higher  than one.

\noindent Another possibility, which is suggested by our results,
would  be to define quantum groups as those which contain symmetries
which are  anomalous under RG transformations. This definition is
independent of the space  dimensionality. The quantum anomalous term
in equation (\ref{11xhh33c}),
and an analogous term also present in our qRG
treatment of the ITF model, gives a discrete  realization of this
idea. A continuum analogue of this anomaly is given by the
Feigin-Fuchs current, which has an anomalous operator product
expansion with the energy-momentum tensor \cite{dotsenko-fateev}. At
this point it may be  worth to recalling  the continuous version of
quantum groups in CFT of reference \cite{g-s}, which uses the
Feigin-Fuchs or free field realization of the  latter. Putting all
these arguments together, we arrive at the conclusion that  quantum
groups are indeed defined by symmetries anomalous under RG
transformations. This point of view about quantum groups may set up
the pathway to new  developments in the field.

Finally, it is somewhat amusing the way in which the word group
enters the title of these notes.
It refers both to the Renormalization
Group method and to Quantum Groups, but neither of them are really
groups!

%(\ref{2})

\vspace{20 pt}

{\bf Acknowledgements}
 \vspace{20 pt}

Work partially supported in part by CICYT under  contracts AEN93-0776
(M.A.M.-D.) and PB92-1092, European Community Grant ERBCHRXCT920069
(G.S.).

%(\ref{2})

% %

\def\baselinestretch{1.5} \noindent  %


\vspace{2cm}









\newpage \begin{thebibliography}{99}

\bibitem{wilson} K.R. Wilson,  {\em  Rev. Mod. Phys.}  {\bf 47}, 773
(1975).

\bibitem{anderson} P.W. Anderson,
{\em J. Phys. C} {\bf 3}, 2436 (1970).

\bibitem{drell} S.D. Drell, M. Weinstein, S. Yankielowicz,
 {\em Phys. Rev.  D} {\bf 16}, 1769 (1977).

\bibitem{jullien} R. Jullien, P. Pfeuty, J.N. Fields, S. Doniach, {\em
Phys. Rev.
 B} {\bf 18}, 3568 (1978).

\bibitem{bpz} A.A. Belavin, A.M. Polyakov and A.B. Zamolodchikov,
{\em   Nucl. Phys.  B}{\bf 241}, 333 ( 1984).


\bibitem {drinfeld-jimbo} V.G. Drinfeld, ``Quantum Groups" in {\em
Proceedings of the 1986 International Congress of Mathematics}, ed.
A.M Gleason (Am.  Math. Soc., Berkeley); M. Jimbo, {\em Lett. Math.
Phys. } {\bf 10}, 63 (1985), {\em  Lett. Math. Phys. } {\bf 11}, 247
(1986).

\bibitem {ags} L. Alvarez-Gaum\'e,  C. G\'omez,  G. Sierra, {\em Phys.
Lett.  B} {\bf 220}, 142 (1989); .{\em   Nucl. Phys. B}{\bf 330}, 347
( 1990).

\bibitem {q-germanyo} M.A. Mart\'{\i}n-Delgado and G. Sierra, {\em
1995,  UCM-CSIC preprint, to be published.}


\bibitem{white-noack} S.R. White, R.M. Noack, {\em Phys. Rev.  Lett.
}{\bf 68}, 3487 (1992).

\bibitem {bc-germanyo} M.A. Mart\'{\i}n-Delgado and G. Sierra, {\em
1995,  UCM-CSIC preprint, to be published in Phys. Lett. {\bf B}.}


\bibitem{white}  S.R. White, {\em Phys. Rev.  Lett.} {\bf 69}, 2863
(1992); {\em Phys. Rev.  B} {\bf 48}, 10345 (1993).



\bibitem {germancesar} C. G\'omez, M. Ruiz-Altaba,  G. Sierra,
``Quantum  Groups in Two-Dimensional Physics". Cambridge University
Press (to be  published).

\bibitem {sklyanin}  E.K. Sklyanin, {\em   J. Phys. A} {\bf 21}, 2375
(  1988).

\bibitem {pasquier-saleur} V. Pasquier, H. Saleur, {\em   Nucl. Phys.
B}{\bf 330}, 523 ( 1990).

\bibitem {jullienlibro}  Pfeuty, P.Jullien, R. and Penson, K.A., in
``Real-Space Renormalization", editors Burkhardt, T.W. and van
Leeuwen, J.M.J., series topics in Current Physics {\bf 30},
Springer-Verlag 1982.

\bibitem {jaitisi}  Gonz\'alez, J.,  Mart\'{\i}n-Delgado, M.A.,
Sierra, G., Vozmediano, A.H., {\em Quantum Electron Liquids and
Hight-$T_c$ Superconductivity}, Lecture Notes in Physics, Monographs
vol. 38, Springer-Verlag 1995.

\bibitem {rabin}J.M. Rabin, {\em Phys. Rev.  B} {\bf 21}, 2027 (1980).


\bibitem {lieb-s-m} Lieb, E., Schultz, T. and Mattis, D. {\em 1961,
Ann. Phys. {\bf 16}, 407}

\bibitem {orbach} Orbach, R., {\em 1958,  Phys. Rev.  {\bf 112}, 309}

\bibitem {alcaraz} F.C. Alcaraz, M.N. Barber, M.T. Batchelor, R.J.
Baxter,  G.R.W. Quispel, {\em   J. Phys. A }{\bf 20}, 6397 (1987).

\bibitem{cardy}  H.W. Blote, J.L. Cardy, M.P. Nightingale,
{\em Phys. Rev.  Lett.} {\bf 56}, 742
(1986).

\bibitem{affleck}  I. Affleck,
{\em Phys. Rev.  Lett.} {\bf 56}, 746
(1986).


\bibitem {gaudin} Gaudin, M., {\em 1971,  Phys. Rev. A {\bf 4}, 386}

\bibitem {cherednik} Cherednik, I.V., {\em 1984,  Theor. Math. Phys.
{\bf 61}, 977}


\bibitem {fpacheco}  A. Fern\'andez-Pacheco, {\em Phys. Rev.  D} {\bf
19}, 3173 (1979).

\bibitem {arnaudon}  D. Arnaudon, A. Chakrabarti, {\em  Comm. Math.
Phys.} {\bf  139}, 461 (1991).


\bibitem {kac-deconcini}  C. de Concini, V.G. Kac, {\em  Prog. Math. }
{\bf  92}, 471 (1990).

\bibitem {jimbo} E. Date, M. Jimbo, K. Miki, T. Miwa, {\em Comm. Math.
Phys. } {\bf  137}, 133 (1991); {\em Publ. Res. Inst. Math. Sci. }
{\bf  27}, 639  (1991).

\bibitem {macoy}  H. Au-Yang, B.M. McCoy, J.H.H. Perk, S. Tan, M.L.
Yan,  {\em   Phys. Lett.  A}{\bf 123}, 219 ( 1987); H. Au-Yang, R.J.
Baxter, J.H.H.  Perk, {\em
  Phys. Lett.  A}{\bf 128}, 138 ( 1988).

\bibitem {nos}  M.A. Mart\'{\i}n-Delgado, G. Sierra, {work in
preparation  .}

\bibitem {hirsch}J.E. Hirsch, {\em Phys. Rev.  B} {\bf 22}, 5259
(1980);  C. Dasgupta, P. Pfeuty, {\em   J. Phys. C }{\bf 14}, 717
(1981); C.  Vanderzande, {\em   Phys. Lett.  A}{\bf 18}, 889 ( 1985);
J. Perez-Conde, P. Pfeuty,  {\em Phys. Rev.  B} {\bf 47}, 856 (1993).


\bibitem {liebwu}  E. Lieb, F. Wu, {\em Phys. Rev.  Lett.} {\bf 20},
1445  (1968); B.S. Shastry, {\em Phys. Rev.  Lett.} {\bf 56}, 2453
(1986).


\bibitem {zamo}  A.B. Zamolodchikov, {\em  Comm. Math. Phys.} {\bf
79},  489 (1981).

\bibitem {dotsenko-fateev} V.S. Dotsenko, V.A. Fateev, {\em   Nucl.
Phys.  B}{\bf },  ( 19).

\bibitem {g-s} C. Gomez, G. Sierra, {\em   Phys. Lett.  B}{\bf 240},
149  ( 1990). {\em   Nucl. Phys. B}{\bf 352}, 791 ( 1991).





\end{thebibliography}
\end{document}